\pdfoutput=1 
\documentclass[doublecol,a4paper]{epl2} 

\usepackage[english,francais]{babel}

\usepackage{amsfonts,amsmath,amssymb,ifpdf,epsfig,graphicx}

\usepackage{amsmath,amssymb,mathrsfs}

\definecolor{M_Beige}         {rgb}{0.96 , 0.96 , 0.86}

\definecolor{M_Brown}         {rgb}{0.65 , 0.16 , 0.16}

\definecolor{M_Gold}          {rgb}{1.00 , 0.84 , 0.00}

\definecolor{M_LemonChiffon}  {rgb}{1.00 , 0.98 , 0.80}

\definecolor{M_Orange}        {rgb}{1.00 , 0.60 , 0.00}

\definecolor{M_Pink}          {rgb}{0.80 , 0.55 , 0.60}

\definecolor{M_Violet}          {rgb}{0.83 , 0.21 , 0.93}

\definecolor{M_Green}          {rgb}{0.2 , 0.6 , 0.2}

\definecolor{M_Gray}          {rgb}{0.4 , 0.4 , 0.4}

\definecolor{M_BluPal}          {rgb}{0.7 , 0.7 , 0.9}

\definecolor{M_DarkBlu}          {rgb}{0 , 0 , 0.75}

\newcommand{\ket}[1]{|\kern.3ex#1\kern.3ex\rangle}
\newcommand{\bra}[1]{\langle\kern.3ex #1 \kern.3ex|}

\newcommand{\mean}[1]{\left\langle #1\right\rangle}
\newcommand{\smean}[1]{\langle #1\rangle}
\newcommand{\EXP}[1]{e^{#1}}         

\newcommand{\im}{\mathop{\mathrm{Im}}\nolimits}      

\def\I{{\rm i}}
\newcommand{\deriv}[2]{\frac{\mathrm{d}#1}{\mathrm{d}#2}}

\def\D{{\rm d}}                  

\def\levy{\mathcal{L}}

\begin{document}

\selectlanguage{english}

\title{Generalized Lyapunov exponent of random matrices and universality classes for SPS in 1D Anderson localisation}

\shorttitle{GLE and universality classes for SPS in 1D Anderson localisation}

\author{Christophe Texier 
}
\shortauthor{C. Texier}

\institute{LPTMS, CNRS, Universit\'e Paris-Saclay, 91405 Orsay cedex, France }



\abstract{ 
 Products of random matrix products of $\mathrm{SL}(2,\mathbb{R})$, corresponding to transfer matrices for the one-dimensional Schr\"odinger equation with a random potential $V$, are studied.
 I consider both the case where the potential has a finite second moment $\langle V^2\rangle<\infty$ and the case where its distribution presents a power law tail $p(V)\sim|V|^{-1-\alpha}$ for $0<\alpha<2$.
  I study the generalized Lyapunov exponent of the random matrix product (i.e. the cumulant generating function of the logarithm of the wave function). 
  In the high energy/weak disorder limit, it is shown to be given by a universal formula controlled by a unique scale (single parameter scaling).
  For $\langle V^2\rangle<\infty$, one recovers Gaussian fluctuations with the variance equal to the mean value: $\gamma_2\simeq\gamma_1$.
  For $\langle V^2\rangle=\infty$, one finds $\gamma_2\simeq(2/\alpha)\,\gamma_1$ and non Gaussian large deviations, related to the universal limiting behaviour of the conductance distribution $W(g)\sim g^{-1+\alpha/2}$ for $g\to0$.
}

\maketitle


The concept of Lyapunov exponent has occupied a central place in the study of random matrix products \cite{BouLac85}, and has found many applications for the physics of disordered systems. Two well-known examples are the random Ising chain, where the Lyapunov exponent coincides with the mean free energy per spin, and that of wave equations with disorder for which it provides a measure of the localisation of the wave \cite{Luc92,CriPalVul93}. 
If one considers the product $\Pi_N=M_N\cdots M_2M_1$ of $N$ independent and identically distributed (i.i.d.) random matrices, the Lyapunov exponent, 
defined as $\tilde\gamma_1=\lim_{N\to\infty}(1/N)\mean{\ln||\Pi_N||}$, 
measures the exponential growth rate of the matrix elements, where $||\cdot||$ is a norm for the matrix ensemble.
Fluctuations are also of importance and can be studied with the \textit{generalized Lyapunov exponent} (GLE) $\widetilde{\Lambda}(q)=\lim_{N\to\infty}(1/N)\ln\mean{||\Pi_N||^q}$, which corresponds to the cumulant generating function for $\ln||\Pi_N||$ (the existence of the limit implies that all cumulants scale as $\sim N$). 
Although the terminology was introduced in \cite{PalVul87}, the concept appeared earlier in the mathematical literature on generalized central limit theorems for non-commutative objects \cite{Tut65,BouLac85}.
Considering the sum $S=\sum_{n=1}^Nx_n$ of i.i.d. \textit{commuting} random variables, distributed according to a distribution $p(x)$, 
it is well-known that the determination of the distribution of the sum 
$P_N(S)$  
in the $N\to\infty$ limit leads to considering different universality classes, depending on the second moment. 
When $\smean{x_n^2}<\infty$, the distribution 
$P_N(S)$ 
is given by the universal Gaussian law (central limit theorem).
When the distribution presents a power law tail, $p(x)\sim|x|^{-1-\alpha}$ for $x\to\pm\infty$, with $0<\alpha<2$, the second moment is infinite $\smean{x_n^2}=\infty$ and the problem belongs to a different universality class, labelled by the exponent $\alpha$.
Then, 
$P_N(S)$ 
is given by a L\'evy law of index $\alpha$, irrespectively of the details of the distribution $p(x)$.
In this letter, it is argued that similar considerations apply to products of random matrices, i.e. \textit{non commuting} objects. 
I consider products of i.i.d. $2\times2$ random matrices of the form
\begin{equation}
  \label{eq:RM}
  M_n = \begin{pmatrix}
    \cos\theta_n & -\sin\theta_n \\\sin\theta_n & \phantom{-}\cos\theta_n
  \end{pmatrix}
  \begin{pmatrix}
    1 & u_n \\ 0 & 1
  \end{pmatrix}
  \:,
\end{equation}
this choice being motivated by the relation with 1D Schr\"odinger equation, for which \eqref{eq:RM} are transfer matrices \cite{ComTexTou10}. 
These physical motivations lead to adopting a slightly different definition of the GLE, involving a fluctuating number of matrices
\begin{equation}
  \label{eq:DefGLE}
  \Lambda(q)=\lim_{x\to\infty} \frac1x  \ln\mean{||\Pi_{\mathscr{N}(x)}||^q} 
  \:,
\end{equation}
where $\mathscr{N}(x)$ is a Poisson process with $\mathrm{Proba}\{\mathscr{N}(x)=N\}=\EXP{-\rho x}(\rho x)^N/N!$.
The GLE \eqref{eq:DefGLE} is the cumulant generating function for $\ln|\psi(x)|\sim\ln||\Pi_{\mathscr{N}(x)}||$, where $\psi(x)$ is the wave function. 
Its determination is an outstanding problem in general, even numerically~\cite{Van10} (apart for integer argument, by making use if the replica trick \cite{Pen82,BouGeoHanLeDMai86,BouGeoLeD86,Van10}). 

\vspace{0.125cm}

\noindent\textbf{Main result.---}
Setting $\theta_n=k\ell_n$ and $u_n=v_n/k$, I discuss here the case where $\ell_n$'s are exponentially distributed, while the coefficients $v_n$'s have an arbitrary distribution $p(v)$.
The case with finite second moment $\mean{v_n^2}<\infty$ ($\alpha=2$) and the case with infinite second moment with power law tail $p(v)\sim|v|^{-1-\alpha}$ for $0<\alpha<2$, are both analysed.
The main result of the paper is the universal form of the GLE
\begin{align}
   \label{eq:TheMainResult}
   \Lambda(q) \simeq&   
   \:\gamma_1 \,
   \frac{2^{\alpha}\Gamma(\frac{1+\alpha}{2})}{\pi^{3/2}\Gamma(\alpha)\Gamma(1+\frac{\alpha}{2})}\,
   \nonumber\\
   &\times
   \Gamma\left(\frac{\alpha+2+q}{2}\right)\, \Gamma\left(\frac{\alpha-q}{2}\right)\,
   \sin\left(\frac{\pi q}{2}\right)
   \:,
\end{align}
for $q\in]-2-\alpha,\alpha[$, obtained in the limit $k\to\infty$.
For $\alpha=2$, Eq.~\eqref{eq:TheMainResult} gives the quadratic behaviour $\Lambda(q)\simeq\gamma_1\,q\,(1+q/2)$.
The fact that the GLE is controlled by a unique scale, the Lyapunov exponent $\gamma_1=\Lambda'(0)$, is known as the \og single parameter scaling \fg{} (SPS) property. 
The present work thus extends SPS to a broad class of disordered models.

\vspace{0.125cm}

\noindent\textbf{Models.---}
I consider here continuous models of localisation: the Schr\"odinger equation for a random potential:
\begin{equation}
  \label{eq:Schrodinger}
  -\psi''(x) + V(x)\,\psi(x) = E\,\psi(x)
  \:.
\end{equation}
The choice of continuous models, rather than discrete lattice models, does not affect the results, as I am interested here in universal properties.
Assuming the absence of spatial correlations for the random potential, one can write
\begin{equation}
  \label{eq:DefLevy}
  \mean{\EXP{-\I s\int_0^x\D t\,V(t)}} = \EXP{-x\,\levy(s)}
  \:,
\end{equation}
where $\levy(s)$ is the L\'evy exponent~\cite{App04}, which will play a central role here
(see also Appendix~1 of Ref.~\cite{GraTexTou14}).  

To make connection with the matrices \eqref{eq:RM}, consider the Frisch-Lloyd model \cite{FriLlo60} corresponding to a random potential of the form $V(x)=\sum_n v_n\,\delta(x-x_n)$, 
where $\ell_n=x_n-x_{n-1}>0$ has a distribution $P(\ell)=\rho\,\EXP{-\rho\ell}$ and the weights an arbitrary distribution $p(v)$ (this corresponds to independent random $\delta$-impurities with mean density $\rho$).
Writing the energy as $E=k^2$, the matrices \eqref{eq:RM} for $\theta_n=k\ell_n$ and $u_n=v_n/k$ are known to be transfer matrices for the vector $\big(\psi'(x_n^-)\,,\,\psi(x_n^-)\big)$ \cite{ComTexTou10,Tex19}, thus $\ln|\psi(x)|\sim\ln||\Pi_{\mathscr{N}(x)}||$. For the Frisch-Lloyd model, the L\'evy exponent has the form 
$\levy(s)=\rho\,\big[1-\hat{p}(s)\big]$, where $\hat{p}(s)=\mean{\EXP{-\I sv_n}}$ is the Fourier transform of $p(v)$ (this expression is more easy to prove by considering the generating functional \cite{GraTexTou14}
$
\smean{\exp\big\{-\I\int\D x\, h(x)\,V(x)\big\}}=\exp\big\{-\int\D x\, \levy(h(x))\big\}
$).
From now on, I assume that $V(x)$ has a symmetric distribution around zero (in particular $\mean{V(x)}=\rho\mean{v_n}=0$), leading to a real symmetric L\'evy exponent, $\levy(-s) = \levy(s)$.
In the universal regime $k\to\infty$, only the $s\to0$ behaviour of the L\'evy exponent is important as I will show.
One has to distinguish two cases
\begin{itemize}
\item  
  The L\'evy exponent has an analytic behaviour for $s\to0$, precisely $\levy(s)\simeq c\,s^2$, where $c$ is some nonuniversal constant (disorder strength).
\item  
  The L\'evy exponent has a non-analytic behaviour at the origin, $\levy(s)\simeq c\,|s|^\alpha$ for $s\to0$, with $\alpha\in]0,2[$.
\end{itemize}
With  $\alpha\in]0,2]$, both situations can be treated on the same footing.
The two cases have a clear interpretation within the Frisch-Lloyd model.
In the first case $c=\rho\,\mean{v_n^2}/2<\infty$, while in the second, the weight distribution presents a power law tail $p(v)\sim|v|^{-1-\alpha}$ leading to $\mean{v_n^2}=\infty$ and $\hat{p}(s)\simeq1-b\,|s|^\alpha$ for $s\to0$. 
\footnote{
  I stress that here, $\alpha$ controls the L\'evy exponent and the case $\alpha=2$ does \textit{not} describe the Frisch-Lloyd model with a tail $p(v)\sim|v|^{-3}$, which corresponds to $\levy(s)\sim-s^2\ln|s|$ for $s\to0$.
}
The strict equality $\levy(s)=c\,|s|^\alpha$ describes the case where $\int_0^x\D t\,V(t)$ is a $\alpha$-stable symmetric L\'evy process \cite{App04} (including the Brownian motion for $\alpha=2$, when $V(x)$ is a Gaussian white noise).

Because the main result \eqref{eq:TheMainResult} fully relies on the $s\to0$ behaviour of the L\'evy exponent, the only crucial assumption for the disordered model is the absence of spatial correlations, leading to the form~\eqref{eq:DefLevy}.

\vspace{0.125cm}

\noindent\textbf{Formalism.---}
The starting point of the present analysis is a result of Ref.~\cite{Tex19}, where the question of fluctuations of random matrix products was addressed in greater generality.
The GLE is the largest eigenvalue of a certain linear operator \cite{Tut65}. 
The spectral problem (Eq.~6.17 of Ref.~\cite{Tex19}) can be formulated as follows:
denote by $\phi(s;\Lambda)$ the solution of the differential equation
\begin{equation}
  \label{eq:TheDiffEq}
  \left[ -\deriv{^2}{s^2} + \frac{q}{s}\deriv{}{s} + E - \frac{\levy(s)+\Lambda}{\I s} \right] 
  \phi(s;\Lambda)  =0
  \:,
\end{equation}
vanishing for $s\to+\infty$.
The solution behaves as  \cite{Tex19}
\begin{equation}
  \phi(s;\Lambda)
  \simeq 1 -\frac{\I\Lambda}{q}\,s +  \omega(\Lambda)\,s^{q+1} + \cdots
  \hspace{0.2cm}\mbox{for }
  s\to0^+
  \:.
\end{equation}
The GLE is the solution $\Lambda=\Lambda(q)$ of the secular equation $\im[\omega(\Lambda)]=0$.

\vspace{0.125cm}

\noindent\textbf{GLE in the universal regime.---}
The solution of the spectral problem is now found in the high energy/weak disorder limit. For $E=k^2\to+\infty$, one expects the GLE to be of the order of the disorder strength, $\Lambda(q)=\mathcal{O}(c)$. The idea is to 
solve the differential equation \eqref{eq:TheDiffEq} by a perturbation method by considering $(\levy(s)+\Lambda)/(\I s)$ as the perturbation.
One writes 
$
  \phi(s;\Lambda) 
  =
  \phi_0(ks)
  +
  \phi_1(ks;\Lambda)
  + 
  \phi_2(ks;\Lambda)
  +\cdots
$
where $\phi_n=\mathcal{O}(c^n)$.
Correspondingly, the coefficient $\omega(\Lambda)$ can be expanded in powers of $c$ as well
$
  \omega(\Lambda) = \omega_0 + \omega_1(\Lambda)+ \omega_2(\Lambda) +\cdots
$.

At order $c^0$ the differential equation 
$-\phi''_0(y)+(q/y)\phi_0'(y)+\phi_0(y)=0$
has solution 
$
   \phi_0(y) = y^{\nu}\,K_{\nu}(y)
$ with $\nu=(q+1)/2$, where $K_\nu(z)$ is the MacDonald function \cite{gragra}.
The order $c^n$ contribution solves 
\begin{equation}
    \left[ 
    - \deriv{^2}{y^2} + \frac{q}{y}\deriv{}{y} + 1\right]
    \phi_n(y;\Lambda)
    = 
    \frac{\levy(y/k)+\Lambda}{\I\, k\,y}\,
    \phi_{n-1}(y;\Lambda)
\end{equation}
Since $\phi_0$ is real, one deduces that $\phi_1\in\I\mathbb{R}$, $\phi_2\in\mathbb{R}$, etc.
Hence $\omega_0\in\mathbb{R}$, $\omega_1\in\I\mathbb{R}$, $\omega_2\in\mathbb{R}$, etc, and the secular equation takes the form 
$
  \omega_1(\Lambda) + \omega_3(\Lambda) +\cdots = 0
$.
In the $E\to+\infty$ limit, one can simply truncate the equation as $\omega_1(\Lambda)\simeq0$.

The solution at order $c^1$ is 
\begin{align}
  \label{eq:phi1}
  \phi_{1}(y;\Lambda)
  &= 
  \frac{\I}{k}\,y^{\nu}
  \bigg\{
    K_{\nu}(y) \int_0^y\D u\, 
    \left(\levy(u/k)+\Lambda\right)\, I_{\nu}(u) \, K_{\nu}(u) 
    \nonumber\\
   &\hspace{0.5cm}
   +  
    I_{\nu}(y) \int_y^\infty\D u\, 
    \left(\levy(u/k)+\Lambda\right)\,  K_{\nu}(u) ^2
  \bigg\}
\end{align}
which vanishes exponentially at infinity, as $\sim\EXP{-y}$.
The problem is now to identify the term $\omega_1(\Lambda)\,y^{q+1}$ for $y\to0$. 

Eq.~\eqref{eq:phi1} makes clear that, in the limit $k\to\infty$, the solution is fully controlled by the $s\to0$ behaviour of the L\'evy exponent $\levy(s)$.
Thus, all the results derived below are completely universal, controlled only by the exponent $\alpha\in]0,2]$ of the L\'evy exponent.

It is easy to see that the first term of \eqref{eq:phi1} only provides contributions $\mathcal{O}(y)$ and $\mathcal{O}(y^{q+2})$, to lowest order in $y$, hence do not contribute to $\omega_1(\Lambda)$.
The leading order of the second term of \eqref{eq:phi1} is easily obtained
\begin{align}
  y^{\nu}
    I_{\nu}(y) 
    &\int_y^\infty\D u\, 
    \left(\frac{\levy(u/k)}{\Lambda}+1\right)\,  K_{\nu}(u) ^2
  \nonumber\\
   &= \frac{2^{\nu}\Gamma(\nu)}{4\nu(2\nu-1)}\, y + \Omega(\Lambda)\,y^{2\nu}+ \cdots
\end{align}
where $\Omega(\Lambda)=-(\I\,k/\Lambda)\omega_1(\Lambda)$.
It is however much more tricky to get the next leading order term 
$\mathcal{O}(y^{2\nu})$, i.e. $\mathcal{O}(y^{q+1})$, and derive the coefficient $\Omega(\Lambda)$.
For $q<\alpha$, the integral $\int_y^\infty\D u\, \levy(u/k)\,  K_{\nu}(u) ^2$ has a limit for $y\to0$. Thus
\begin{align}
  \label{eq:BigOmega}
\Omega(\Lambda) 
=&
\frac{1}{2^\nu\Gamma(\nu+1)}
\bigg[
\frac{c}{\Lambda\,k^\alpha}\int_0^\infty\D u\,u^\alpha\,K_\nu(u)^2
  \\\nonumber
&+
\lim_{y\to0}
\left\{
  \int_y^\infty\D u\,K_\nu(u)^2 - 2^{2\nu-2}\Gamma(\nu)^2\,\frac{y^{-2\nu+1}}{2\nu-1}
\right\}
\bigg]
\end{align}
One uses (formula 6.576 of \cite{gragra})
\begin{align}
  \label{eq:EqFrom6.576Grad}
   &\int_0^\infty\D u\,u^\alpha\,K_\nu(u)^2
  \\\nonumber
   =& \frac{2^{\alpha-2}}{\Gamma(1+\alpha)}\,
   \Gamma\left(\frac{1+\alpha}{2}\right)^2\,
   \Gamma\left(\frac{1+\alpha}{2}+\nu\right)\,
   \Gamma\left(\frac{1+\alpha}{2}-\nu\right)
\end{align}
for $\alpha>2|\nu|-1$, i.e. $-2-\alpha<q<\alpha$, 
and
\begin{align}
  \label{eq:ALimit}
  \lim_{y\to0}\left\{
  \int_y^\infty\D u\,K_\nu(u)^2 - 2^{2\nu-2}\Gamma(\nu)^2\,\frac{y^{-2\nu+1}}{2\nu-1}
  \right\}
  =\frac{\pi^2}{4\cos\pi\nu}
\end{align}
for $0<\nu<3/2$, i.e. $-1<q<2$ [for $0<\nu<1/2$, the integral converges for $y=0$ and the equation is simply given by setting $\alpha=0$ in \eqref{eq:EqFrom6.576Grad}].
From (\ref{eq:BigOmega},\ref{eq:EqFrom6.576Grad},\ref{eq:ALimit}), one sees that the secular equation $\Omega(\Lambda)\simeq0$ gives
\begin{align}
   \label{eq:TheMainResultText}
   \Lambda(q) \simeq   
   \frac{c\,k^{-\alpha}\Gamma(\frac{1+\alpha}{2})}{\pi^{3/2}\Gamma(1+\frac{\alpha}{2})}\,
   \Gamma\!\left(\frac{\alpha+2+q}{2}\right) \Gamma\!\left(\frac{\alpha-q}{2}\right)
   \sin\!\left(\frac{\pi q}{2}\right)
\end{align}
for $q\in]-2-\alpha,\alpha[$. 
One checks the symmetry relation  \cite{Van10,Tex19}
\begin{equation}
\Lambda(q)=\Lambda(-2-q)
\:.
\end{equation}

The normal case for disorder with finite second moment corresponds to $\alpha=2$:
from \eqref{eq:TheMainResultText}, one gets $\Lambda(q)\simeq\big[c/(4k^2)\big]\,q\,(1+q/2)$. 
Thus the cumulants $\gamma_n$ with $n>2$ are subleading in the disorder; the same conclusion was obtained by studying the first cumulants for a specific continuous model \cite{SchTit02} and a lattice model \cite{SchTit03,TitSch03} (the quadratic form corresponds to the Gaussian distribution derived first in Refs.~\cite{AntPasSly81,AltPri89}). 

\vspace{0.125cm}

\noindent\textbf{Cumulants and single parameter scaling.---}
The GLE is a cumulant generating function.
Writing 
\begin{equation}
  \label{eq:ExpansionGLE}
  \Lambda(q)=\sum_{n=1}^\infty\frac{\gamma_n}{n!}\,q^n
  \:,
\end{equation}
one identifies $\gamma_nx$ as the cumulant of order $n$ of $\ln|\psi(x)|\sim\ln||\Pi_{\mathscr{N}(x)}||$.
In the past, a lot of effort has been devoted to derive the first cumulants for various continuous and discrete models, motivated to prove the \og \textit{single parameter scaling} \fg{} (SPS) hypothesis within microscopic models. 
SPS hypothesis, a corner stone of localisation theory, states that the scaling properties of disordered systems are controlled by a unique parameter~\cite{AbrAndLicRam79}. This property is expected to hold for the full distribution of the wave function~\cite{AndThoAbrFis80,Sha86}.
When the second moment of the disordered potential is finite, $\mean{V^2}<\infty$, the distribution of $\ln|\psi(x)|$ is Gaussian \cite{AntPasSly81,AltPri89} and SPS takes the form
\begin{equation}
  \label{eq:SPS}
  \gamma_2 \simeq \gamma_1
  \:.
\end{equation}
Equivalently, for a weakly conducting sample of long lentgh $L$,
the dimensionless conductance is $g\sim|\psi(L)|^{-2}$ from Borland's conjecture \cite{Bor63}, and SPS is $\mathrm{Var}(\ln g)\simeq -2\mean{\ln g}\simeq4\gamma_1L$ \cite{CohRotSha88}. 
It was later recognized that the relation \eqref{eq:SPS} is only valid asymptotically in the weak disorder/high energy regime, where universality is expected \cite{CohRotSha88} (cf.~\cite{SchSchSed04} for a more formal discussion).
The interest for SPS in 1D disordered systems was renewed in 2000 with the search of a universal criterion separating SPS and non-SPS regimes \cite{DeyLisAlt00}. 
\footnote{the criterion proposed in these papers was ruled out by a counter example in Ref.~\cite{Tex19}.} 
The authors of this reference have supported their analysis by an exact calculation of the variance for the Lloyd model, a tight binding model with on-site energies distributed according to a Cauchy law \cite{Llo69}.
Besides its theoretical interest, the Lloyd model was shown to be relevant in various contexts: 
for the quantum kicked rotor model of dynamical localisation \cite{FisGrePra82} or, more recently, in disordered ladders where the Cauchy disorder arise effectively due to the presence of flat bands \cite{Luc19}.
The popularity of the Lloyd model comes from the possibility to get several analytical results like the Lyapunov exponent \cite{Tho72,Ish73,Luc92}, its variance \cite{DeyLisAlt00} 
or higher cumulants \cite{TitSch03}.
In particular, for the Lloyd model, the relation between the two cumulants was shown to present an additional factor of two, $\gamma_2\simeq2\,\gamma_1$ \cite{DeyLisAlt00} 
instead of \eqref{eq:SPS}, 
whose origin was never explained, to the best of my knowledge. 
The framework presented here provides a simple interpretation. The expansion of \eqref{eq:TheMainResult} gives the variance
\begin{equation}
  \label{eq:BeautifulSPS}
   \gamma_2    \simeq   \frac{2}{\alpha}\,\gamma_1 
  \hspace{0.5cm}\mbox{for }
  \alpha\in]0,2]
  \:,
\end{equation}
which extends~\eqref{eq:SPS} corresponding to $\alpha=2$. 
For the Cauchy case ($\alpha=1$), the relation $\gamma_2\simeq2\gamma_1$ of \cite{DeyLisAlt00,TitSch03} has now a clear interpretation in terms of the exponent $\alpha$.

Moreover, Eq.~\eqref{eq:TheMainResultText} is of the form $ \Lambda(q) \simeq(\alpha+q)\,f_\mathrm{odd}(q)$, with $f_\mathrm{odd}(-q)=-f_\mathrm{odd}(q)$. This implies 
\begin{equation}
  \label{eq:WonderfulSPS}
 \gamma_n \simeq \frac{n}{\alpha}\, \gamma_{n-1} \hspace{0.5cm}\mbox{for } n \mbox{ even,}
\end{equation}
further generalizing \eqref{eq:BeautifulSPS} 
(for $n>2$, Eq.~\eqref{eq:WonderfulSPS} holds for $\alpha<2$).
Thus even cumulants are deduced from odd ones.
Expansion of \eqref{eq:TheMainResult} in powers of $q$ gives the first cumulants 
\begin{align}
  &\gamma_1 \simeq \frac{c\,\Gamma(\alpha) }{(2k)^\alpha}
  \:,\hspace{0.5cm}
  \gamma_3 \simeq \left[\frac{3}{2}\psi'\left(\frac{\alpha}{2}\right)-\frac{\pi^2}{4}\right] \gamma_1 
  \:, 
  \\
 &\gamma_5 \simeq
   \left[
      \frac{5}{8}\psi'''\left(\frac{\alpha}{2}\right)
    + \frac{15}{4}\psi'\left(\frac{\alpha}{2}\right)^2
    - \frac{5\pi^2}{4}\psi'\left(\frac{\alpha}{2}\right)
    +\frac{\pi^4}{16}
    \right] \gamma_1  
    \:,
    \nonumber
\end{align}
etc,
where $\psi(z)$ is the digamma function.
For $\alpha=2$, one recovers the well-known perturbative result $\gamma_1\simeq c/(4E)$ (see \cite{AntPasSly81,LifGrePas88}).
For $\alpha<2$, the anomalous energy dependence $\gamma_1\sim k^{-\alpha}$ was obtained earlier in Ref.~\cite{BieTex08}. 
Rescaling the GLE \eqref{eq:TheMainResultText} by $\gamma_1$ leads to the universal form \eqref{eq:TheMainResult}.

\vspace{0.125cm}

\noindent\textbf{Cauchy disorder.---}
The case of Cauchy disorder ($\alpha=1$) has been much studied \cite{Tho72,Ish73,LifGrePas88,Luc92,DeyLisAlt00,TitSch03} 
and deserves a special discussion. 
From \eqref{eq:TheMainResultText}, one gets the rescaled GLE
\begin{equation}
  L(q)
  =  \!\!
  \lim_{E/c\to\infty} \!\!
  \frac{\Lambda(q)}{\gamma_1}=\frac2\pi(q+1)\tan\left(\frac{\pi q}{2}\right)
  \hspace{0.25cm}\mbox{for }
  q\in]-3,1[
  \:.
\end{equation}
One can write $L(q)=\sum_{n=1}^\infty(\kappa_n/n!)\,q^n$, where $\kappa_n=\lim_{E\to+\infty}\gamma_n/\gamma_1$ are the rescaled cumulants, given by 
\begin{equation}
   \kappa_n = 4 \pi^{n-2}\left( 2^n -1 \right) |B_n|
   \hspace{0.5cm}\mbox{and}\hspace{0.5cm}
   \kappa_{n-1} = \frac{\kappa_n}{n}
\end{equation} 
for $n$ even, where $B_n$'s are the Bernoulli numbers \cite{gragra}.
In particular $\kappa_1 =1$, $\kappa_2 = 2$, $\kappa_3 =\pi^2/2\simeq4.93$, $\kappa_4=2\pi^2\simeq19.7$, $\kappa_5=\pi^4\simeq97.4$, $\kappa_6=6\pi^4$, etc.  
This perfectly agrees with the estimation given in \cite{TitSch03} within a lattice model  (the paper gave $\kappa_3\simeq5$, 
\footnote{the equality ``$\kappa_3=5$'' in \cite{TitSch03} is probably a typo.}
$\kappa_4\simeq20$ and $\kappa_5\simeq100$).  
The greater efficiency of the method presented here, compared to the approach of Refs.~\cite{DeyLisAlt00,SchTit03,TitSch03}, 
is due to the different strategies~:
in these latter works, the first coefficients of the expansion \eqref{eq:ExpansionGLE} were studied, considering the $E\to\infty$ limit on each cumulant.
Here, the strategy was to consider first the high energy limit of the GLE, which has allowed to obtain systematic expressions of the cumulants by expanding in powers of $q$ afterwards.

\vspace{0.125cm}

\noindent\textbf{Large deviations and conductance distribution.---}
For $\alpha<2$, the GLE \eqref{eq:TheMainResultText} 
diverges as 
$\Lambda(q)\sim1/(\alpha-q)$ for $q\to\alpha^-$ (and diverges symmetrically for $q\to(-\alpha-2)^+$). 
Correspondingly, the distribution of the modulus of the wave function $|\psi(x)|$ presents power law large deviation tails
\begin{align}
  \mathscr{P}_x(\psi)\sim
  \begin{cases}
  \psi^{1+\alpha}  &\mbox{for }  \psi\to0
  \\
  \psi^{-1-\alpha}  &\mbox{for }  \psi\to\infty
  \end{cases}
\end{align}
while typical fluctuations are described by the log-normal behaviour
\begin{equation}
  \mathscr{P}_x(\psi)\sim
  \frac1\psi \exp\left\{- \frac{\alpha}{4\gamma_1x}\,( \ln \psi - \gamma_1x)^2\right\}
  \:.
\end{equation}
The tail for $\psi\to\infty$ corresponds to weakly conducting samples, with dimensionless conductance $g\sim|\psi(L)|^{-2}$, where $L$ is the sample length.
Hence, the distribution of the conductance presents the universal large deviation tail
\begin{equation}
  \label{eq:DistribConductance}
   W_L(g)\sim g^{-1+\alpha/2}
  \mbox{ for }g\to0
  \:.
\end{equation}
In \cite{TitSch03}, a power law behaviour was identified, although the exponent was not obtained.
The result \eqref{eq:DistribConductance} is in agreement with the numerics of Ref.~\cite{MenMarGopVar16}.

\vspace{0.125cm}

\noindent\textbf{Universality.---}
I have based the discussion on the Schr\"odinger equation for a potential characterized by a L\'evy exponent $\levy(s)$, i.e. with no correlation in space. This includes the Frisch-Lloyd model.
I have stressed above that the main result \eqref{eq:TheMainResult} only relies on the $s\to0$ behaviour of $\levy(s)$, and not on the details of the disorder distribution.
Another model much studied is the Kronig-Penney model with a regular lattice of $\delta$-impurities of random weights. Writing the potential as 
$V(x)=a^{1/\alpha}\sum_n\eta_n\delta(x-na)$, with $\alpha=2$ for $\mean{\eta_n^2}<\infty$ and $0<\alpha<2$ for $p(\eta\to\pm\infty)\sim|\eta|^{-1-\alpha}$, 
one obtains easily the generating functional
$ \smean{\EXP{-\I\int\D x\,h(x)\,V(x)}} \simeq \exp\{-c\,a\sum_n|h(na)|^\alpha\} $. 
Considering $a\to0$ or equivalently when $h(x)$ is smooth on the scale $a$, the sum can be replaced by an integral:
this shows that the large scale properties of the Kronig-Penney model can also be described with the formalism of the letter for L\'evy exponent $\levy(s\to0)\simeq c|s|^\alpha$.

Finally, I stress that the analytical expressions obtained here for $\alpha=1$ coincides with the estimations of the first cumulants obtained in Refs.~\cite{DeyLisAlt00,TitSch03} for Lloyd's lattice model, which emphasizes further the universal character of the results presented in the letter.

\vspace{0.125cm}

\noindent\textbf{Conclusion.---}
In this paper, I have derived the generalized Lyapunov exponent for certain products of random matrices of $\mathrm{SL}(2,\mathbb{R})$ characterizing wave function statistics in the 1D Schr\"odinger equation. 
A broad variety of disordered models was considered, which has allowed me to derive a universal form for the GLE in the high energy/weak disorder limit, Eq.~\eqref{eq:TheMainResult}.
For disorder with finite second moment ($\alpha=2$), one has $\Lambda(q)\simeq\gamma_1\,q\,(1+q/2)$ for $q\in]-4,+2[$, which characterizes Gaussian fluctuations. 
Out of this interval, the behaviour of the GLE was however not discussed above.
For the model with Gaussian white noise potential, $\levy(s)=c\,s^2$, the \textit{non universal} behaviour 
$\Lambda(q)\sim c^{1/3}\,|q|^{4/3}$ for $q\to\pm\infty$ was derived in \cite{BouGeoLeD86,FyoLeDRosTex18} (see also \cite{Tex19}). 
\footnote{
  This behaviour was obtained for $q>0$ and $E=0$ in Refs.~\cite{BouGeoLeD86,FyoLeDRosTex18}.
  When $E\neq0$, it holds for $|q|\gg E^{3/2}/c$.
} 
%
%
For disorder with power law distribution (exponent $\alpha\in]0,2[$), the universal expression \eqref{eq:TheMainResult} was derived for $q\in]-2-\alpha,\alpha[$, with the GLE diverging at the boundaries of the interval.
Interestingly, this shows that, for small but \textit{finite} disorder, universality is stronger for power law disorder with $\mean{V^2}=\infty$ than in the more standard case with $\mean{V^2}<\infty$. 
In this latter case, the large deviations are non universal (dominated by higher cumulants $\gamma_n$ with $n>2$ subleading in the disorder strength), while in the former all cumulants scale the same way and the GLE is universal over its whole interval of definition.

An interesting open question is to investigate the universality of the GLE for models within other symmetry classes: 
in the presence of a chiral symmetry, disordered models are known to present ``anomalies'' which have been widely studied. The first cumulants have been determined for a lattice model in \cite{SchTit03} (see also \cite{RamTex14} for a study of $\gamma_2$ within a continuous model), but the GLE is still unknown in this case.
As Ref.~\cite{RamTex14} makes clear, this problem is also related to the challenging determination of the GLE for other types of random matrix products. A starting point could be to investigate random matrix products in the continuum limit \cite{ComLucTexTou13,Tex19}.

Finally, a more challenging issue would be extend the results to the multichannel case, or higher dimensions relevant for the problem of Ref.~\cite{FyoLeDRosTex18}.

\vspace{0.125cm}

\noindent\textbf{Acknowledgments.---}
I thank Alain Comtet and Yves Tourigny for numerous stimulating discussions (other aspects of the work published in the paper \cite{Tex19}, on which the present letter is based, are exposed in Ref.~\cite{ComTexTou19}).


\end{document}